\documentstyle[aps,preprint,epsfig]{revtex}
\tighten

\begin{document}

\def\mev{\hbox{\ MeV}}
 
\draft

%% Include following two lines for Journal Style

 %\twocolumn[\hsize\textwidth\columnwidth\hsize  %**Journal
 %\csname @twocolumnfalse\endcsname              %**Journal

\title{Exceptional points and double poles of the $S$ matrix}

\author{I.~Rotter}
\address{
Max-Planck-Institut f\"ur Physik komplexer Systeme,
D-01187 Dresden, Germany}

\date{\today}

\maketitle

\vspace*{.5cm}

\begin{abstract}
Exceptional points and
double poles of the $S$ matrix  are both characterized
by the  coalescence of a pair of eigenvalues. In the 
first case, the coalescence causes a defect of the Hilbert space. 
In the second case, this is not so as  shown in prevoius  papers.
Mathematically, the reason for this difference is the bi-orthogonality 
of the eigenfunctions of a non-Hermitian operator that is 
ignored in the first case. The consequences 
for the topological structure of the Hilbert space
are studied and compared with existing experimental data.

\end{abstract}

\vspace*{0.2cm}

\section{Introduction}

Information on 
the topological structure of the Hilbert space 
can be obtained  from a study of its singularities.
Berry \cite{berry} showed that  geometric
phases appear when a diabolic point is surrounded 
by varying adiabatically external parameters of a quantum system.
Manifestations of this phase factor have been considered
and  proven experimentally already in the eighties
in many different fields of physics, e.g. \cite{proof1}.
In 1994, it has been studied also by using microwave resonators: 
the sign change of the wave function has been found
after a cyclic excursion around a diabolic point in the space of 
shapes of the resonator  \cite{lauber}.

Other singularities are  exceptional points \cite {kato}
which appear in the complex $\Lambda$ plane of the eigenvalues  
${\cal E}_k(\Lambda)$ of the Hamiltonian $H=H_0 + \Lambda H_1$. Their 
positions are characteristic of the Hamiltonian $H$, once 
$H_0$  and $H_1$ are given \cite{heiss,hemuro,mondr} (which both are 
assumed to be real and symmetric). 
The exceptional points are characterized by the 
coalescence of a pair of eigenvalues, i.e. ${\cal E}_l(\Lambda_{\rm EP})
= {\cal E}_k(\Lambda_{\rm EP})$.  When the corresponding eigenfunctions
are assumed to be orthogonalized in the standard manner, it follows
 $\psi_l(\Lambda_{\rm EP}) = \psi_k(\Lambda_{\rm EP})$.  
This means, that $\psi_k(\Lambda_{\rm EP})$ can not be normalized 
at $\Lambda = \Lambda_{\rm EP}$, since the orthogonality conflicts 
with the normalization requirement. As a consequence, 
an exceptional point is 
characterized by the fact that the rank of the associated matrix 
$H_0 + \Lambda_{\rm EP} H_1$ drops by one
at $\Lambda = \Lambda_{\rm EP}$ and the
two wave functions coalesce into one. This implies a 
defect of the underlying Hilbert space \cite{kato}.     

In \cite{demb}, the topological structure of exceptional points is
studied experimentally by using a microwave resonator. The exceptional
point   is surrounded 
by varying adiabatically external parameters of the system.
As a result, the eigenvalues and eigenvectors are exchanged while 
encircling an exceptional point, but one of the eigenvectors undergoes
a sign change which can be discerned in the field patterns. 
From these results, the authors draw the conclusion that the 
exceptional points can clearly be distinguished from other topological
singularities such as diabolic points.

In describing physical processes,  the exceptional points
lead to problems. First, the splitting of the Hamiltonian $H$ 
into $H_0$ and $H_1$ can not be done arbitrarily. For a fixed
$H_0$, the part $H_1$ is well defined since it describes 
the coupling of the states
of the system (described by $H_0$) via the environment (continuum 
of decay channels) into which it is embedded \cite{ro91}. 
Secondly, the property
${\cal E}_l(\Lambda_{\rm EP}) = {\cal E}_k(\Lambda_{\rm EP})$
 is characteristic of a double pole of the $S$ matrix. Here, 
 ${\cal E}_k = E_k - i/2 \Gamma_k $ is the complex energy of the 
 resonance state $k$ with energy $E_k$ and width $\Gamma_k$.
The $S$ matrix describes physical processes, and no 
hints at all to  defects of the Hilbert space are known
at  a double pole. For numerical examples see 
the results of calculations performed in a schematical model 
\cite{mudiisro} and for atoms \cite{kylstra,marost}.

Theoretical studies have shown that the topological 
structure of avoided level crossings is directly related to the
topological structure of double poles of the $S$ matrix 
being branch points in the complex plane
\cite{ro01}. The transition from a double pole of the
$S$ matrix to an avoided level crossing by varying a parameter 
occurs continuously. The avoided level crossings are directly
related to the diabolic points \cite{berry}. Thus, the topological
structure of a double pole of the $S$ matrix and that of a 
diabolic point are  related to one another. 

The problem is now the following.
The double pole of the $S$ matrix and the exceptional point 
are both characterized  by the
coalescence of two eigenvalues of a non-Hermitian Hamilton
operator at a certain value of a parameter. 
Nevertheless, their topological structures are
different: according to \cite{demb},
the topological structure of an exceptional point differs from that
of a diabolic point while the topological structure of the 
double pole is related to that of the diabolic point, 
as discussed above \cite{ro01}.
The question arises therefore what differences exist
between the exceptional points and the double poles of the $S$ matrix
that could cause  their different topological structures.

In order to find an answer to this question, 
the Schr\"odinger equation has to be solved in the whole function space 
containing everything, i.e. discrete and continuous states.
By using a projection operator technique, an effective Hamiltonian
can be derived from this Schr\"odinger equation 
which describes the system ($Q$ subspace) after 
embedding it into the continuum of decay channels
($P$ subspace) \cite{ro91}.
Its eigenvalues and eigenfunctions are complex. 
The eigenvalues  coincide with the poles of the $S$ matrix.
The eigenfunctions are related to the
wave functions of the resonance states by
a Lippmann-Schwinger-like relation \cite{rosmatr}. They
are bi-orthogonal.   At the double pole of the $S$
matrix,   the Hilbert space has no defect
due to the bi-orthogonality of the wave functions.  The $S$ matrix 
behaves smoothly by varying  parameters also when  the double pole 
is met \cite{ro01}.

It is the aim of the present paper to derive the phase changes of the
wave functions that appear after surrounding a double pole of 
the $S$ matrix parametrically. In Sect. II, the relation between
the eigenvalues of the effective Hamilton operator and the poles 
of the $S$ matrix is scetched while in Sect. III the relation between
double poles  and avoided level crossings  is discussed. 
The double poles of the $S$ matrix are branch points in the complex plane. 
At these points,  the wave functions
of the two states are exchanged, $\psi_1 \to \pm \, i\, \psi_{2}$.
This causes a mixing of the wave functions
in the region of  avoided level crossings. In Sect. IV,
 the phase changes appearing after surrounding a diabolic point 
 and a double
pole of the $S$ matrix are derived and compared with experimental results. 
They agree with  all data for isolated crossings of two states that 
are published in \cite{lauber,demb}. Conclusions
on the topological structure of the function space are drawn in the last 
section.

\section{Effective Hamiltonian and poles of the $S$ matrix}

After embedding a system into the continuum of decay channels, the 
discrete states of the system turn over in resonance states  with
a finite lifetime. The Hamiltonian of the system becomes effectively 
non-Hermitian with complex eigenvalues
${\cal E}_k = E_k - i/2 \Gamma_k$, where the width $\Gamma_k$ is
inverse proportional to the lifetime. 

The relation between the poles of the $S$ matrix and the complex
eigenvalues ${\cal E}_k$ can be derived from the Schr\"odinger equation
\begin{eqnarray}
(H - E)\Psi_E = 0
\label{bas1}
\end{eqnarray}
with the  Hamilton operator $H$ and the set $\{ \Psi_E \}$ of
wave functions containing the discrete states of the system as well as 
the scattering wave functions of the environment into which the system
is embedded. The operator $H$ is hermitian. 

In a first step, two sets of equations have to be  solved:
\begin{eqnarray}
(H^{\rm cl} - E_k^{\rm cl})\, \Phi_k^{\rm cl} = 0
\label{bas2}
\end{eqnarray}
and
\begin{eqnarray}
\sum_{c'}(H^{cc'} - E ) \, \xi_E^{c'(+)} = 0 
\label{bas3}
\end{eqnarray}
where $H^{\rm cl}$ describes the system with the discrete states $k$ 
and $H^{cc'}$ the continuum with coupled decay channels $c$. Then,
the two projection operators are defined by
\begin{eqnarray}
Q = \sum_k |\Phi_k^{\rm cl}\rangle \langle \Phi_k^{\rm cl}|
\qquad P = \sum_c \int {\rm d}E \, |\xi_E^{c\, (+)} \rangle 
\langle \xi_E^{c\, (+)}|
\label{bas4}
\end{eqnarray}
and $H^{\rm cl}$ is identified with $QHQ \equiv H_{QQ}$ and 
$H^{cc'}$ with $PHP \equiv H_{PP}$. The two other terms of 
$H = H_{QQ} + H_{PP} + H_{QP} + H_{PQ} $ describe the 
coupling between the two subspaces. The solutions of coupled
channel equations with source term
\begin{eqnarray}
\sum_{c'}(H^{cc'} - E )\, \langle \xi_E^{c'(+)}|\, \omega_k \rangle =  
-\, \langle \xi_E^{c\, (+)} |H_{PQ}|\Phi_k^{\rm cl} \rangle
\label{bas5}
\end{eqnarray}
provide the wave functions $\omega_k$ that contain the coupling 
between the two subspaces.

Using the completeness relation $P+Q=1$, one obtains for the solution
of the whole problem \cite{ro91}
\begin{eqnarray} 
\Psi_E^c
   =  \xi_E^{c(+)} + \frac{1}{\sqrt{2 \pi}} \sum_{k=1}^N
  {\tilde \Omega}_k    \cdot
  \frac{
\tilde\gamma_{k}^{c}}
  {E - {\tilde E}_k + \frac{i}{2} {\tilde \Gamma}_k} \; .  
\label{bas6}
\end{eqnarray}
Here, 
\begin{eqnarray}
\tilde \Omega_k = 
 \tilde\Phi_k + \tilde\omega_k = 
(1 + G_P^{(+)}  H_{PQ})\; \tilde\Phi_k 
\label{bas7}
\end{eqnarray}
is the wave function of the resonance state $k$,
$G_P^{(+)}=P(E-H_{PP})^{-1}P$ is the Green function in the $P$ subspace,
$\tilde \omega_k$ is determined by (\ref{bas5}) with 
$\Phi_k^{\rm cl}$ replaced by $\tilde\Phi_k$, and
\begin{eqnarray}
\tilde \gamma_{k}^{c}(E) = \sqrt{2\pi} \; 
\langle \tilde \Phi_k^*  | H_{QP} | \xi_{E}^{c\, (+)} \rangle
= \sqrt{2\pi} \; 
\langle \xi_{E}^{c\, (+)} | H_{PQ} |\tilde \Phi_k \rangle 
\; .
\label{bas11}
\end{eqnarray}
Further, $\tilde\Phi_k$ is  eigenfunction and  $\tilde{\cal E}_k
 = \tilde E_k - i/2 \;\tilde\Gamma_k$  eigenvalue of the effective 
Hamiltonian
 \begin{eqnarray}
{\cal H} = H_{QQ} + H_{QP} G_P^{(+)} H_{PQ}
\label{bas8}
\end{eqnarray}
that describes the system after embedding it into the continuum of 
decay channels. ${\cal H}$ is non-hermitian, its eigenvalues and 
eigenvectors are complex. The eigenfunctions are bi-orthogonal,
\begin{eqnarray}
\langle \tilde \Phi_k^* | \tilde \Phi_l \rangle = \delta_{k,l} 
\label{bas9}
\end{eqnarray}
where $\tilde \Phi_k^{\rm right} \equiv \tilde \Phi_k  $
and $\tilde \Phi_k^{\rm left} =  \tilde \Phi_k^* $ \cite{ro01,pegoro1}. 
As a consequence
\begin{eqnarray}
\langle \tilde \Phi_k | \tilde \Phi_l \rangle
={\rm Re} ( \langle \tilde\Phi_k|\tilde\Phi_k \rangle) 
  & ; & \quad  A_k \equiv   \langle \tilde \Phi_k|\tilde\Phi_k \rangle
\ge 1             
\nonumber \\
 \langle \tilde\Phi_k|\tilde\Phi_{l\ne k} \rangle  =  
i \; {\rm Im} (\langle \tilde\Phi_k|\tilde\Phi_{l\ne k} \rangle )  
 =  -\langle \tilde\Phi_{l\ne k}|\tilde\Phi_k \rangle 
  & ; & \quad   B_k^{l\ne k} \equiv  
| \langle\tilde \Phi_k|\tilde\Phi_{l\ne k} \rangle| \ge 0 \; . 
 \label{bas10} 
\end{eqnarray}

Using (\ref{bas6}), (\ref{bas7})  and the 
Lippmann-Schwinger equation for the scattering wave functions,
one gets for the resonance part of the
$S$ matrix \cite{ro91,ro01}
\begin{eqnarray}  
 S_{cc'}^{(\rm res)} =  i \; \sum_{k=1}^N
  \frac{\tilde\gamma_{k}^{c}\; \tilde\gamma_{k}^{c'}}
  {E - {\tilde E}_k + \frac{i}{2} {\tilde \Gamma}_k} \; .
\label{bas12}
\end{eqnarray}
The $\tilde \gamma_{k}^{c}$ are
the coupling matrix elements of the resonance states to the 
continuum. $S_{cc'}^{\rm (res)} $ describes the 
resonance part of the $S$ matrix also in the overlapping regime. 
The interferences 
between the resonance states are taken into account by diagonalizing the 
effective Hamiltonian ${\cal H}$. Due to the unitarity of the $S$ matrix, 
the  $\tilde \gamma_k^c, \; \tilde E_k$ and $\tilde \Gamma_k$ are energy 
dependent functions. The relation $\tilde \Gamma_k = 
 \sum (\tilde \gamma_k^c)^2 $ holds only for isolated resonances.
In the overlapping regime,
the energy dependence of both functions is different, as a rule. For
numerical examples see \cite{rep02}. 

As can be seen from  (\ref{bas12}), the poles of the $S$ matrix are 
determined by the eigenvalues $\tilde{\cal E}_k$ of the effective Hamiltonian
(\ref{bas8}) after solving the fixed-point equations 
${\cal E}_k=\tilde{\cal E}_k(E=E_k) $ \cite{ro01}.
As an example, resonances of a microwave cavity are studied experimentally
in the overlapping regime \cite{stm}. The results show the phenomenon
of resonance trapping  and are described well by  
(\ref{bas12}) with the effective Hamiltonian (\ref{bas8}).

\section{Double poles of the $S$ matrix and avoided level crossings}

The relation between double poles of the $S$ matrix and avoided level
crossings can be illustrated best by means of a simple two-level model.
Let us consider the complex two-by-two Hamiltonian matrix
 \begin{eqnarray}
{\cal H} =
 \left(
\begin{array}{cc}
 e_1(\lambda) - \frac{i}{2}\gamma_1 & \omega \\
\omega  &   e_2(\lambda) -  \frac{i}{2} \gamma_2
\end{array}
\right) 
\label{4}
\end{eqnarray}
where $e_k$ and  $\gamma_k$ ($k=1,2$) are the unperturbed energies  
and widths, respectively, of the two states. The $e_k$ are assumed to 
depend on the parameter $\lambda$  in such a manner that  
the two states may cross in energy at $\lambda^{\rm cr}$ when $\omega = 0$.
The two states interact only via $\omega $ which is assumed 
in the following  to be 
independent of the parameter $\lambda$ (as the $ \gamma_k$).
The eigenvalues of  ${\cal H}$     are 
\begin{eqnarray}
E_\pm - \frac{i}{2}\,  \Gamma_\pm  = 
\frac{1}{2}\;\bigg[(e_1 + e_2) - \frac{i}{2} \; 
(\gamma_1   +  \gamma_2 )\bigg] \, \pm \frac{1}{2} \;
\sqrt{F} 
\label{4a}
\end{eqnarray}
with
\begin{eqnarray}
 F=
\bigg[( e_1 - e_2) - \frac{i}{2} \; (\gamma_1  -  \gamma_2 )\bigg]^2 
+ 4 \, \omega^2 \; .
\label{5}
\end{eqnarray}
When $ F(\lambda,\omega) = 0$ at $\lambda=\lambda^{\rm cr}$ 
(and $\omega=\omega^{\rm cr}$),
the $S$ matrix has a double pole.  

According to  (\ref{5}), $ F = F_R + i \, F_I$ is generally 
a complex number. For illustration, let us discuss the  case 
with real  $\omega$. Then
$\; e_1 = e_2$ at $\lambda=\lambda^{\rm cr}$  and 
we have to differentiate between three cases
\begin{eqnarray}
  F_R(\lambda,\omega) > 0 & \quad \to \quad  & \sqrt{F_R} = {\rm real} 
\label{eq:fri11}\\
  F_R(\lambda,\omega) = 0 & \to  &\sqrt{F_R} = 0 
\label{eq:fri12}\\
  F_R(\lambda,\omega) < 0 &  \to  &\sqrt{F_R} = {\rm imag} \; .
\label{eq:fri13}
\end{eqnarray}
The first case gives the  avoided level
crossing in energy with an exchange of the two wave functions 
at $\lambda^{\rm cr}$. The second case 
corresponds to the double pole of the $S$
matrix. In the third case, the two levels cross freely in energy
and the two states
are {\it not} exchanged at the critical value  
$\lambda^{\rm cr}$ \cite{ro01}.
In \cite{brent}, the
two cases $F_R > 0$ and  $F_R < 0$ are studied experimentally in
a microwave cavity and called {\it overcritical} and
{\it subcritical} coupling, respectively.
The more complicated cases with complex $\omega$ 
are considered in \cite{ro02}. 

The example with real $\omega$ illustrates nicely the relation 
between a double pole of the $S$ matrix and  avoided
or even free crossings of two levels in the complex plane.  
The double pole is a branch point in the complex plane.
The number of these branch points
is of measure zero, but their influence on the 
dynamics of quantum systems can be traced in the many avoided 
level crossings. While the wave functions 
of the two states are exchanged just at the
double pole of the $S$ matrix and are unmixed at any value
of the parameter different from the critical one, this is not so
at an avoided level crossing. In this case, the wave functions
remain mixed in
a certain range of the parameter around the critical value.
This fact has a strong influence on the mixing of all the wave
functions of a system when the level density is high, and
different  avoided level crossings appear at 
values of the parameter inside
this range. For the results of numerical studies see \cite{ro01}.  
 
 The bi-orthogonality relation (\ref{bas9}) holds everywhere, including
at the double pole of the $S$ matrix. The reason is that $A_k \to \infty;
\; B_k^l \to \infty $ [Eq. (\ref{bas10})] and that  
$\langle \tilde \Phi_k^* | 
\tilde \Phi_l \rangle $ is the difference between two infinitely large
numbers (but not their sum). 
This difference may be $0$ (for $l \ne k$) or $1$ (for $l = k$). Thus, 
the orthogonality and normalization requirements do not conflict with
one another and the Hilbert space has no defect at all.
For the results of numerical studies see \cite{ro01}.

It should be mentioned here, that the bi-orthogonality 
of the $\{\tilde \Phi_k\}$
follows directly from the non-Hermiticity of ${\cal H}$. Only for
the eigenfunctions of a Hermitian operator holds 
$\tilde\Phi_k^{\rm left} = \tilde\Phi_k^{\rm right}$
\cite{pegoro1}. Due to the symmetry of ${\cal H}$ it 
holds $\tilde\Phi_k^{\rm left} = \tilde\Phi_k^{\rm right \; *}$ 
for its eigenfunctions  what results in  (\ref{bas9}) for
the bi-orthogonality relation.

Further analytical studies \cite{ro01} have shown that 
the wave functions of the two states at the double
pole of the $S$ matrix are exchanged. It is
\begin{eqnarray}
\tilde \Phi_k^{\rm bp} \to \pm \, i \, \tilde \Phi_{l\ne k}^{\rm bp}
\label{bp1}
\end{eqnarray}
in approaching  the double pole of the $S$ matrix. 
This result is confirmed by numerical studies on 
laser induced continuum structures in atoms  \cite{marost}.

The real and imaginary parts of the wave functions of two resonance states  
as a function of an external parameter increase limitless in approaching
the double pole of the $S$ matrix \cite{ro01}. The sign of  the 
imaginary  part jumps at the double pole (when 
$\tilde\Phi_1 \to +\, i \,\tilde \Phi_2$).  When the double 
pole is not met by varying the external parameter, but the levels avoid 
crossing at the critical value of the parameter, the real and imaginary 
parts remain finite but the jump of the sign remains. 
The wave function 
\begin{eqnarray}
\tilde \Phi_{\rm ch} = a_1 \, \tilde\Phi_1 \pm i \, a_{2} \, 
\tilde\Phi_{2}
\; .
\label{bp2}
\end{eqnarray}
changes  smoothly (without any jump of the sign of its components)
for   $a_2 \to a_1$
at the double pole of the $S$ matrix or at the critical
value of the parameter where the levels avoid  crossing.
For the results of a numerical study see \cite{ro01}.

The diabolic points are related to avoided  crossings
of discrete levels. They occur 
by varying two independent parameters: at the diabolic point, two 
energy surfaces drawn over the plane of the two external parameters 
touch each other at one point  forming a double cone.

\section{Geometric phases}

Let us now consider the geometric phases appearing after encircling 
a diabolic point and a branch point in the complex plane
(double pole of the $S$ matrix), respectively.
In any case, the paths of encircling are 
characterized by the value  $F$, Eq. (\ref{5}), which vanishes only
at the branch point
in the complex plane. Most interesting are states whose 
eigenvalues are near to the real axis. We can restrict our
discussion therefore to real $\omega$ (see Sect. III). 

For encircling the diabolic point or the branch point in the 
complex plane, two external parameters have to be varied.  In the 
experiment \cite{lauber}, the diabolic point is surrounded by varying 
the shape of the microwave resonator by means of two parameters
but leaving the coupling strength to the antenna unchanged. Since 
the two levels considered avoid crossing, the whole path of encircling 
the diabolic point is in the overcritical regime. That means,
the critical value of the parameter is passed
twice, on the way forth as well as back,  under overcritical conditions, 
and the wave functions are exchanged each time when
the critical value of the parameter is reached. 
This is not so in the experiment \cite{demb}
where one of the two parameters is the coupling strength 
of the cavity to another one. Therefore, the
critical value of the parameter is passed on the  path 
of encircling the exceptional point (or branch point in the complex 
plane) only once under overcritical conditions. The other part
of the path is in the subcritical regime where 
the wave functions are nowhere  exchanged (see Sect. III).

In detail:
The diabolic point is surrounded in the experiment \cite{lauber} 
in the regime of overcritical coupling along the whole way of 
encircling and  $\lambda^{\rm cr}$ is passed twice in opposite directions,
(i) $\tilde\Phi_k \to - i \, \tilde \Phi_l; \; \tilde \Phi_l \to + i \, \tilde
\Phi_k$, i.e.
\begin{eqnarray}
\{\tilde\Phi_1, \; \tilde\Phi_2\} \; \to \; \{- \, i \, \tilde\Phi_2, \;  
 + \, i \, \tilde\Phi_1 \}
\label{phas1}
\end{eqnarray}
 and (ii),  on the way back,
$\tilde\Phi_l \to - i\, \tilde \Phi_k; \; \tilde \Phi_k \to + i\, \tilde
\Phi_l$, i.e.
\begin{eqnarray}
\{-i \, \tilde\Phi_2, \;  + i \, \tilde\Phi_1 \} \; \to \; \{ 
- \tilde\Phi_1, \;-  \tilde\Phi_2 \} \; .
\label{phas2} 
\end{eqnarray}
The  phase change occuring after one surrounding the diabolic point
is therefore
\begin{eqnarray}
\{ \tilde\Phi_1, \; \tilde\Phi_2 \}  \; \to \; \{ 
- \,  \tilde\Phi_1, \;  - \,  \tilde\Phi_2 \} \; .
\label{phas2a} 
\end{eqnarray}
This corresponds to the geometric phase discussed by Berry \cite{berry}.

The way of encircling  the branch point in the complex plane 
itself passes from
a region with  overcritical coupling at $\lambda^{\rm cr}$ to another one with 
subcritical coupling at $\lambda^{\rm cr}$. An exchange of the
wave functions takes place only at overcritical coupling
where the resonances avoid crossing.
Thus, a first full surrounding gives
\begin{eqnarray}
\{\tilde\Phi_1, \;   \tilde\Phi_2 \} \; \to \; \{ 
- \, i \, \tilde\Phi_2, \;  + \, i \, \tilde\Phi_1 \}
\label{phas3}
\end{eqnarray}
and a second one (in the same direction)
\begin{eqnarray}
\{ - \, i \, \tilde\Phi_2, \;  + i \, \tilde\Phi_1 \} \; \to \; \{
 + \tilde\Phi_1, \; + \tilde\Phi_2 \} \; .
\label{phas4}
\end{eqnarray}
That means, surrounding the branch point in the complex plane 
twice restores   the wave functions
$\tilde\Phi_k$ including their phases.
This corresponds to the  result
obtained for  surrounding  the diabolic point twice.
In both cases, the wave functions including their phases are restored 
after a second encircling in the same direction:
\begin{eqnarray}
\{ \tilde\Phi_1, \; \tilde\Phi_2 \} \; \Rightarrow \; \{
  \tilde\Phi_1, \;  \tilde\Phi_2 \} \; .
\label{phas5a} 
\end{eqnarray}
 
Encircling the branch point in the complex plane in the opposite direction
gives 
\begin{eqnarray}
\{\tilde\Phi_1 , \; \tilde\Phi_2 \} \; \to \; \{ 
+ \, i \, \tilde\Phi_2  , \; - \, i \, \tilde\Phi_1 \} \; .
\label{phas5}
\end{eqnarray}
Since the experiment \cite{demb} is not sensitive to the possible 
occurence of  a phase $i$ of the wave function, the results
(\ref{phas3})   for one loop with a certain orientation
of the path  and 
(\ref{phas5}) with the opposite  orientation of the path 
agree with the experimental data given in \cite{demb}.
There are no experimental data in \cite{demb}   for the 
phase changes after a second loop. 

An experimental study of interferences
between atomic levels in a laser field is expected \cite{ro02} to 
allow conclusions on the phase changes, including those  
after a second loop.

\section{Concluding remarks}

In the present paper, the phase changes occuring after encircling
parametrically 
an isolated diabolic point and an  double pole of the $S$ matrix
(branch point in the complex plane) are calculated. The results 
are shown to agree with  all  experimental data 
that are published in  \cite{lauber,demb}.
 
The results of \cite{lauber} point to the interesting fact that
the phase changes after surrounding higher-order degeneracies
are more complicated than those obtained 
after encircling a diabolic point. This result has given rise to
further theoretical studies, e.g. \cite{pisto}.  

The experimental results \cite{demb} are interpreted by the authors 
on the basis of exceptional points. This interpretation  leads to the 
conclusion that an exceptional point can  clearly be
distinguished from other topological singularities such as diabolic 
points. The authors claim the following: 
encircling the exceptional point a second time completely with the
same orientation one obtains $\{- \tilde \Phi_k, - \tilde \Phi_l \}$ 
while the next complete loop yields $\{- \tilde \Phi_l,  \tilde \Phi_k \}$ 
and only the fourth loop restores fully the original pair 
$\{\tilde \Phi_k, \tilde \Phi_l \}$. The authors show  experimental 
results only for one complete loop.  No data are given for two or 
more loops. 

The appearance of a phase change of both wave functions
after a second loop around the exceptional point,
$\tilde\Phi_k \Rightarrow  - \tilde\Phi_k; \;
\tilde\Phi_l \Rightarrow   - \tilde\Phi_l $,
suggested in \cite{demb},  does not agree with 
the result  (\ref{phas5a}) obtained for a second complete loop around 
a branch point in the complex plane. According to (\ref{phas5a}), the 
original pair $\{\tilde \Phi_k,  \tilde \Phi_l \}$  is restored 
already after a second complete loop when it is
completed with the same orientation. 
This result coincides with that obtained for a second loop around
a diabolic point. 
It is an expression for the fact that diabolic points 
and branch points in the complex plane are related to one 
another as discussed in this paper.

The results for one loop can not differentiate
between the two interpretations  since the experiment is not 
sensitive to the possible occurence of a phase $i$ in the wave 
function. It can therefore not  be concluded from the
published experimental data whether or not the  topological 
structure studied in \cite{demb} is  different from
that of a diabolic point. Further experimental studies are 
necessary, maybe on atoms in a laser field as suggested in \cite{ro02}.

\vspace{1cm}

{\bf Acknowledgment:}
I am indebted to M. Lewenstein, member of the Editorial Board 
of Physical Review Letters, for the 
suggestion to write the present paper instead of a Comment on \cite{demb}.

\end{document}